\documentclass[a4paper,12pt]{article}
\usepackage{amsmath}
\usepackage{gensymb}
\usepackage{graphicx}
\usepackage{relsize}
\usepackage{upgreek}
\usepackage{hyperref}

\begin{document}
\title{Silicon nitride bent asymmetric coupled waveguides with partial Euler bends}
\author{P. Chamorro-Posada\\
  Dpto. de Teor\'{\i}a de la Se\~nal y Comunicaciones\\
 e Ingenier\'{\i}a Telem\'atica,\\
 Universidad de Valladolid, ETSI Telecomunicaci\'on,\\
 Paseo Bel\'en 15, 47011 Valladolid, Spain}

\maketitle
\begin{abstract}

  Waveguide geometries combining bent asymmetric coupled structures and adiabatic transitions shaped as partial Euler bends are put forward and theoretically analyzed in this work.  Designs aiming to reduce the radiation loss, with application in curved waveguide sections and  Q-enhanced microresonators, and polarization selection geometries, both for the silicon nitride platform, are studied using highly accurate numerical techniques.
\end{abstract}

\section{Introduction}

Photonic integrated circuits (PICs) were initially designed primarily to meet the needs of optical fiber data transmission systems \cite{zhang}.  In recent years, we have witnessed a burst in photonic integration technologies, with an ever-growing range of applications.  Highly active fields include optical sensors \cite{wuJ}, medical applications \cite{wang}, optical frequency comb generation \cite{chang}, and quantum technologies \cite{luo}, to name just a few.  The constant progress in integrated photonic technologies has been led by the development of a large ecosystem, including foundries that provide open access fabrication services \cite{siew}.  Silicon photonics, based on highly mature CMOS fabrication processes, plays a prominent role in this scenario~\cite{siew}.  Even though the traditional silicon on insulator (SOI) technology is still dominant within the CMOS platforms, PICs based on silicon nitride waveguides are particularly appealing for certain applications \cite{sharma}.  When compared with silicon guiding structures, those fabricated with silicon nitride provide smaller linear and nonlinear intrinsic propagation losses, a lower thermo-optic coefficient, and a far wider transparency region that opens the platform for applications ranging from the visible to the mid-infrared.  On the negative side, the main drawback of silicon nitride stems from the fact that its refractive index is smaller than that of silicon.  Therefore, the field confinement in silicon nitride waveguides is poorer and the radiation losses in curved waveguide sections become larger \cite{radiation}.  This ultimately limits the minimum acceptable radii of curvature in integrated devices and, consequently, the scale of integration.

Radiation loss in bent integrated waveguides can be reduced by modifying the waveguide geometry through incorporating subwavelength gratings \cite{wu} or side grooves \cite{harjanne,sole}.  Nevertheless, these design strategies require additional non-standard fabrication steps.  The use of matched bends \cite{melloni2003} permits mitigating the losses at the transitions between a constant curvature section and the straight input and output waveguides by adjusting the total extent of the bend to a multiple of the beat length of its first two modes. An alternative method, which can be applied to bend sections of arbitrary length, is to maximize the mode coupling at the discontinuity by applying a relative lateral shift to the straight and bent waveguides \cite{neumann,kitoh}.  Other schemes are based on the progressive modification of the bent waveguide width \cite{ladouceur,yi,song,song2020} or shaping the bend using trigonometric \cite{liu}, spline \cite{harjanne,mustieles,bogaerts}, Euler \cite{cherchi,jiang,fujisawa,vogelbacher}, Bezier \cite{yi,sun}, or $n$-adjustable \cite{zhang2023} functions.  Bend radiation losses can also be minimized using different algorithms \cite{chen2012,gabrielli,yu2018,liu2018,bahadori,li2020,yu2020}. 

Asymmetric coupled geometries can be effectively used to tailor the optical field emission in bent integrated curved waveguides \cite{radiation}.  A careful design of such geometries can substantially reduce the radiation loss, thus permitting diminished footprints in microresonators \cite{Q,Qexp} and in-circuit interconnects.  These guiding structures can also be employed for the design of improved performance microring label-free biosensors \cite{biosensors}.  In addition, they provide high differential losses for the TE and TM polarizations that can be exploited for the implementation of ultracompact, all-dielectric, broadband integrated polarizers with  high polarization extinction ratios and low insertion losses \cite{POL,POLexp} in a silicon nitride platform.

The aforementioned applications are based on trimming the radiation properties in the curved two-waveguide sections but, generally, they also require a mechanism designed to deal with the discontinuities at the straight--bent waveguide sections.  In previous works \mbox{\cite{Q,Qexp,POL,POLexp}}, the lateral offset technique \cite{neumann,kitoh} was used for this purpose.  Actually defining the abrupt transitions demanded by lateral offsets requires resolutions beyond the standard fabrication methods.  When these geometries are used in photolithography masks, de facto, they are smoothed by the fabrication process, resulting in adiabatic transitions \cite{Qexp,POLexp}.  Even though fabricated devices have evidenced performances close to those associated with ideal conditions, reliable designs based on feasible geometries are desirable.  In this work, we address the use of Euler bends in combination with asymmetric coupled waveguides of constant curvature.  Such partial Euler bends were put forward by Fujisawa et al. for the silicon photonic platform \cite{fujisawa} and later reformulated by Vogelbacher et al. for the silicon nitride platform \cite{vogelbacher}.  Whereas in these works the footprint of the whole bent section is kept constant, we analyze designs with a fixed radius of the constant curvature section.  Therefore, the impact of the extent of the partial Euler bent section on the device footprint has to be analyzed separately.

\section{Device Geometry}

\begin{figure}[h]
  \begin{tabular}{cc}
    \includegraphics[width=5 cm]{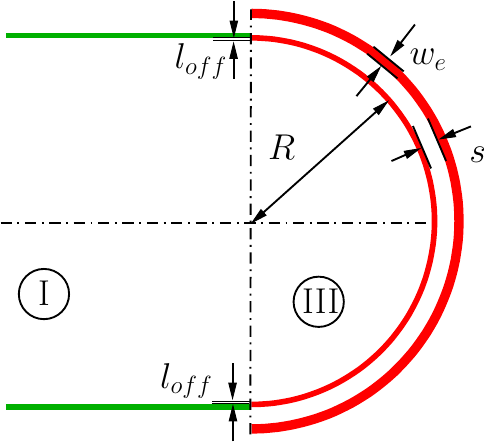}&
    \includegraphics[width=6.5 cm]{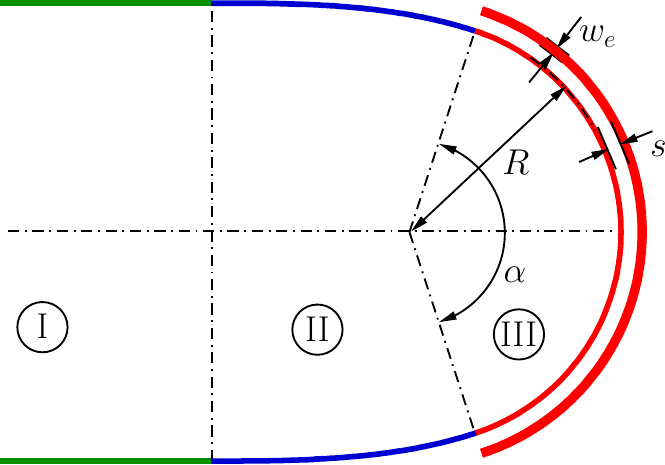}
    \end{tabular}
\caption{(\textbf{Left}) Conventional geometry for the implementation of bent asymmetric coupled waveguides based on lateral offset. (\textbf{Right}) Target geometry based on partial Euler bends analyzed in this work. \label{fig::geometry}}
\end{figure}

The device geometry addressed in this analysis is displayed in Figure \ref{fig::geometry}.   Depending on the target design, this construction could be part of a Q-enhanced racetrack micro-resonator as in \cite{Q,Qexp}, a polarizer \cite{POL,POLexp}, or any other circuit bent section.  Even though the bend angle was set to 180$^\circ$, the results could be extrapolated to other values. In the case of polarizer design, several sections like the one displayed in Figure \ref{fig::geometry} can be concatenated to achieve the required device performance \cite{POL,POLexp}.  The cross-section of the main waveguide is assumed to be equal in regions I.
, II and III of Figure \ref{fig::geometry}.  In section I, we find the straight input and output waveguides.  In the calculations, we inject the input field into the lower waveguide and we read the output at the upper waveguide.  Section III is the asymmetric coupled bend section.  There, the guiding structure has a constant curvature, where the radius $R$ is measured to the center of the main waveguide.  The external waveguide has a width $w_e$ and there is a separation $s$ between the external and internal edges of the main and wrapping waveguides, respectively.     

Section II in Figure \ref{fig::geometry} is a clothoid \cite{fujisawa,vogelbacher} with linearly varying curvature from the two extreme values, $\kappa=0$ and $\kappa=1/R$, matching the constant values at sections I and III, respectively.  Parameter $p$ sets the angular extent of the constant curvature section III to a value $\alpha=(1-p)\pi$ and, therefore, it also specifies the actual length of the clothoid of section II.  $p$ can take any value between 0 and 1.   For $p=0$, section II vanishes and, for $p=1$, section III fades out, resulting in a full Euler bend geometry.   The impact on device performance of this gradually varying curvature section as a replacement for the lateral offset \cite{neumann,kitoh} previously used in \cite{Q,Qexp,POL,POLexp} is analyzed in this work through extensive numerical simulations.

For a given curvature radius $R$ and partial section portion $p$, the position along the lower partial Euler bend of region II in Figure \ref{fig::geometry}, relative to the end of the straight input waveguide, in parametric form, reads
\begin{equation}
  \begin{split}
    \Delta x(\zeta)&=2^{2/3}R\sqrt{\pi p}\,\mathlarger{\cal C}\left(\sqrt{\dfrac{p\pi}{2}}\zeta\right)\\
     \Delta y(\zeta)&=2^{2/3}R\sqrt{\pi p}\,\mathlarger{\cal S}\left(\sqrt{\dfrac{p\pi}{2}}\zeta\right),
  \end{split}
\end{equation}
where $\zeta$ varies in the range $\left[0,1\right]$, and $\mathlarger{\cal C}\left(z\right)=\int_0^z\cos(t^2)dt$ and $\mathlarger{\cal S}\left(z\right)=\int_0^z\sin(t^2)dt$ are the Fresnel integrals.  For their evaluation, we used the first four terms of the series \cite{abramowitz}
\begin{equation}
  \begin{split}
    \mathlarger{\cal C}\left(z\right)&=\sum_{n=0}^\infty\left(-1\right)^n\dfrac{z^{4n+1}}{\left(2n\right)!(4n+1)}\\
     \mathlarger{\cal S}\left(z\right)&=\sum_{n=0}^\infty\left(-1\right)^n\dfrac{z^{4n+3}}{\left(2n+1\right)!(4n+3)}
    \end{split}
\end{equation}

The simulated structures consisted of deep etched Si\textsubscript{3}N\textsubscript{4} rectangular waveguides and SiO\textsubscript{2} cladding with the same parameters as in \cite{POLexp}.  In the subsequent discussion, we will refer to the inner waveguide in section III of Figure \ref{fig::geometry} as the {main waveguide}.  The width of the main waveguide was $w=1$ $\upmu$m
, and all waveguides had a height of \mbox{$h=0.3$ $\upmu$m}.  The operation wavelength was $\lambda=1.55$ $\upmu$m.  The refractive index of silicon nitride was $n_{Si\textsubscript{3}N\textsubscript{4}}=1.9835$ and that of the oxide $n_{SiO\textsubscript{2}}=1.4456$.  

\section{Modal Properties of Constant-Curvature  Asymmetric Two-Waveguide Sections}
\begin{figure}[h]
 \begin{tabular}{cc}
   \includegraphics[width=6.5cm]{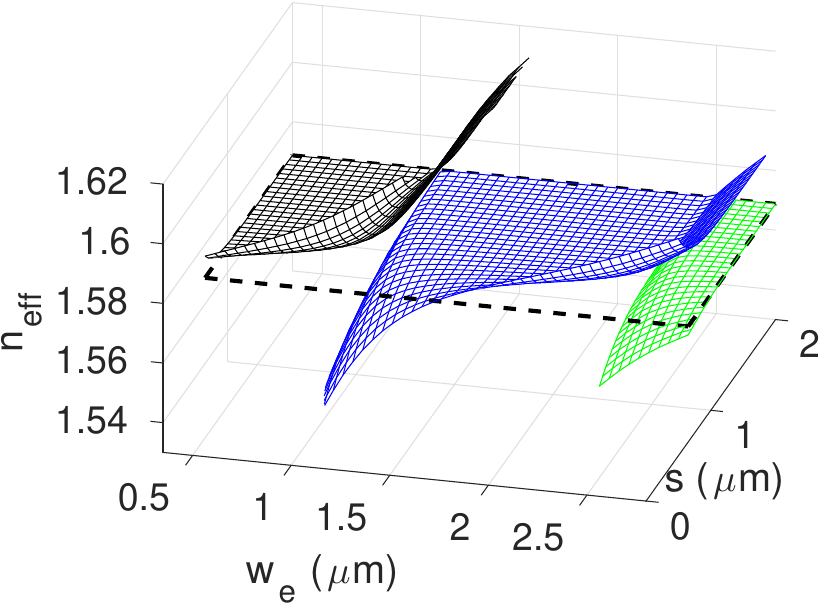}&
      \includegraphics[width=6.5cm]{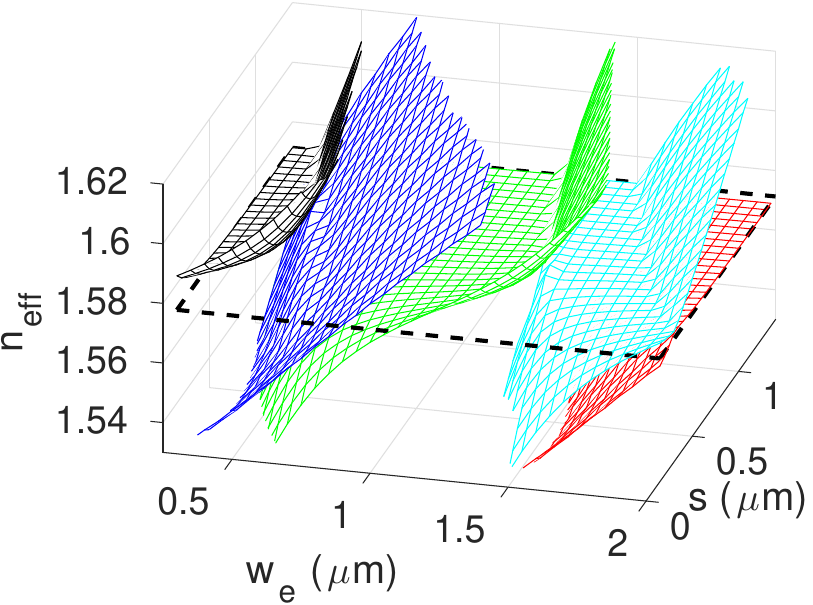}
   \end{tabular}
\caption{(\textbf{Left}) 
 Effective indices of the TE modes in a straight asymmetric coupler (region III of Figure~\ref{fig::geometry} for $R\to\infty$) as a function of $w_e$ and $s$. Sheets corresponding to differents modes are plotted with distinct colors. (\textbf{Right}) Real part of the effective indices of the quasi-modes in the curved asymmetric coupler depicted as region III in Figure~\ref{fig::geometry} for $R=25$ $\upmu$m.} \label{fig::hojas}
\end{figure}

\begin{figure}[h]
  \begin{tabular}{cc}
  \includegraphics[width=6.5 cm]{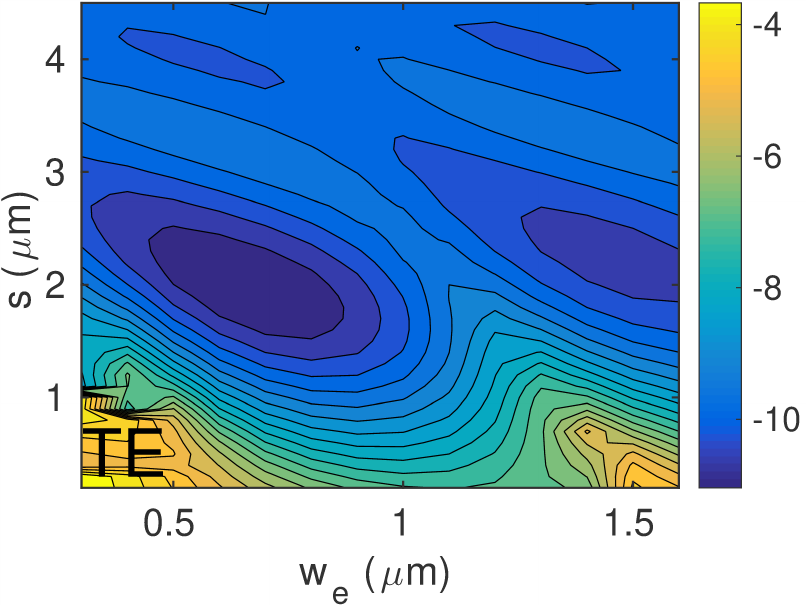}&
  \includegraphics[width=6.5 cm]{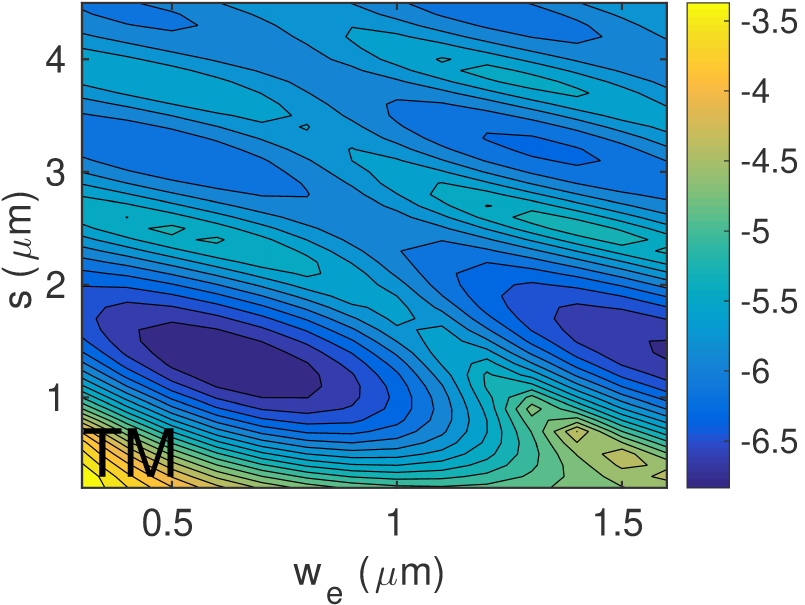}
  \end{tabular}
\caption{{(\textbf{Left}) } 
logarithm of the imaginary part of the effective index of the TE principal mode in the curved asymmetric coupler depicted as region III in Figure~\ref{fig::geometry} for $R=25$ $\upmu$m. (\textbf{Right}) Corresponding results for the TM polarized principal mode. }\label{fig::tetm}
\end{figure}

The modal properties of the guiding structure in section III of Figure \ref{fig::geometry} that determine the physical principles of operation are depicted in Figures \ref{fig::hojas} and \ref{fig::tetm}. The calculation of the data collected in these figures was performed using the \emph{wgms3d} v2.0 {software package} 
 \cite{krause,wgms3d}.

The left plot in Figure \ref{fig::hojas} shows the effective indices of the modes guided by the structure of section III in the straight, $\kappa=0$, case as a function of $w_e$ and $s$.  The total number of supported modes grows with the value of $w_e$ but, for each duple ($w_e$,$s$), we can select from the set of propagating modes one that is predominantly confined within the main waveguide.  We will refer to this particular mode as the {{principal mode}} and it is determined as the one that has an effective index closest to that of the isolated main waveguide; that is, without the coupled external guide.  The constant value of the effective index of the only mode guided in the main waveguide without the coupled waveguide corresponds to the plane outlined in Figure~\ref{fig::hojas} with a dashed line.  When ($w_e$,$s$) is such that there is one modal sheet very close to the plane defining the propagation of the mode of the main waveguide, the corresponding principal mode will be very close to it and the asymmetric coupled structure will evidence a very good field confinement of the principal mode in the main waveguide.  For the $\kappa=0$ asymmetric coupled waveguides, the boundaries defining the principal mode in  ($w_e$,$s$) space are straight lines at constant values of $w_e$.

When asymmetric coupled waveguides are bent, the curvature has the effect of converting the guided modes of the straight geometry into quasi-modes with complex effective indices, where the imaginary part is linked to the radiation loss \cite{radiation,hiremath}.  The right plot of Figure \ref{fig::hojas} illustrates the effects produced by the bending of asymmetric coupled waveguides on the real part of the effective index of the quasi modes for a particular value of $R=25$ $\upmu$m.  First, there is a shift in the real part of the effective index of the mode in the isolated curved main waveguide.  This value is also depicted as a plane outlined by a dashed contour in the right plot of Figure \ref{fig::hojas}. The discussion in the former paragraph is still valid and, for each ($w_e$,$s$), one can define a principal mode that will show the best confinement in the main waveguide, as was performed for the straight waveguides.  On the other hand, the curvature has the effect of distorting the shape of the domains defining the modal sheet that corresponds to the principal mode.

Figure \ref{fig::tetm} displays the effect of bending on the imaginary part of the refractive index.  In those regions of the ($w_e$,$s$) parameter space where the modal sheet of the principal mode is very close to the plane outlined by the dashed contour in Figure \ref{fig::hojas}, the field confinement is very good and the outer waveguide acts as a shield for the radiation field \cite{radiation}.  The net result is that the imaginary effective index attains values below those of the mode of the isolated main waveguide. As shown in the plots of Figure \ref{fig::tetm}, these conditions can be obtained both for TE and TM fields.  A bent asymmetric coupled section designed to fulfill this condition will permit reducing the radiation loss for a given polarization \cite{Q,Qexp}.   On the other hand, when ($w_e$,$s$) is set close to the boundaries corresponding to the jumps of the modal sheet of the principal mode, the mode confinement is poor and radiation losses are increased over those of the isolated main waveguides.  In this case, the external waveguide has the effect of enhancing the radiation.  Figure \ref{fig::tetm} also illustrates the fact that the low/high radiation loss domains in ($w_e$,$s$) space do not overlap for TE and TM fields.  Therefore, it is possible to define values of  ($w_e$,$s$) for which one polarization is selectively attenuated \cite{POL,POLexp}.   

\section{Results}

The 180$^\circ$ bend arrangements depicted in Figure \ref{fig::geometry} describe the layouts chosen for the performance evaluation of the geometries addressed in this work.  By including straight input and output waveguide sections, we could simultaneously take into account radiation losses at the straight--bend transitions and those associated with waveguide curvatures. The evaluation of the transmission properties was carried out using the eigenmode expansion  (EME) method implemented in FIMMPROP (Opt05) \cite{fimmprop}.  This algorithm does not have the limitations of the beam propagation method (BPM) for the analysis of high index contrast waveguides \cite{huang,lui,radiation} and provides highly accurate calculations, with a numerical load significantly lower than other alternatives such as the finite differences in the time domain (FDTD) method \cite{fdtd}.  The optimization of the constant curvature sections was based on modal analyses of curved guiding structures.  These calculations were performed using the {FIMMWAVE} version 7.4.0 (x64)  
 software package.  Both FIMMWAVE and FIMMPROP are commercial software from Photon Design.

\subsection{Reduction in Radiation Loss at Small Radii}\label{sec::loss}

We first address the performance of the curved geometries based on asymmetric coupled waveguides designed to minimize the transmission loss. We assumed designs targeted to TE polarized fields and we sought reduced radii of curvature with diminished radiation losses that would permit reducing the footprint of PIC components.  Example applications are Q-enhanced microresonators \cite{Q,Qexp} or simply curved waveguide sections in the PIC outline.  

\begin{figure}[h]
  \begin{tabular}{cc}
  \includegraphics[width=6.5 cm]{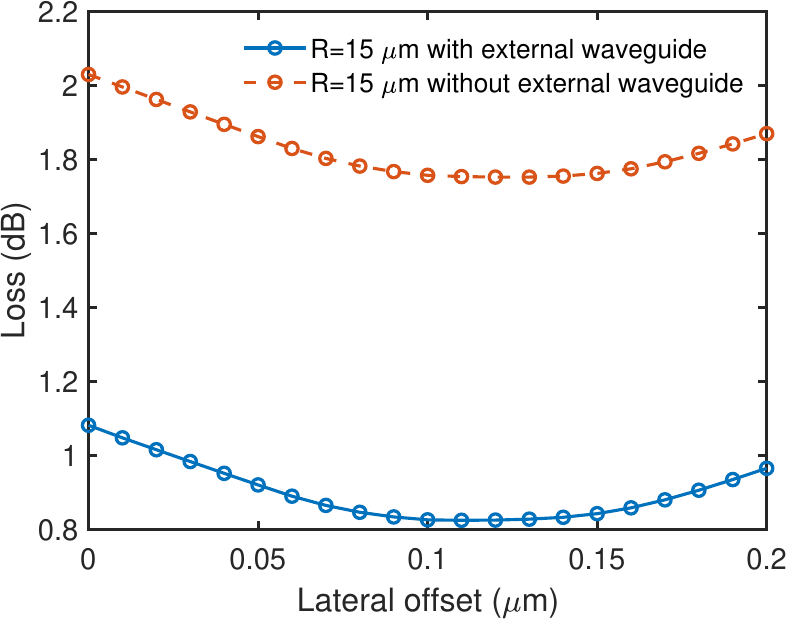}&
  \includegraphics[width=6.5 cm]{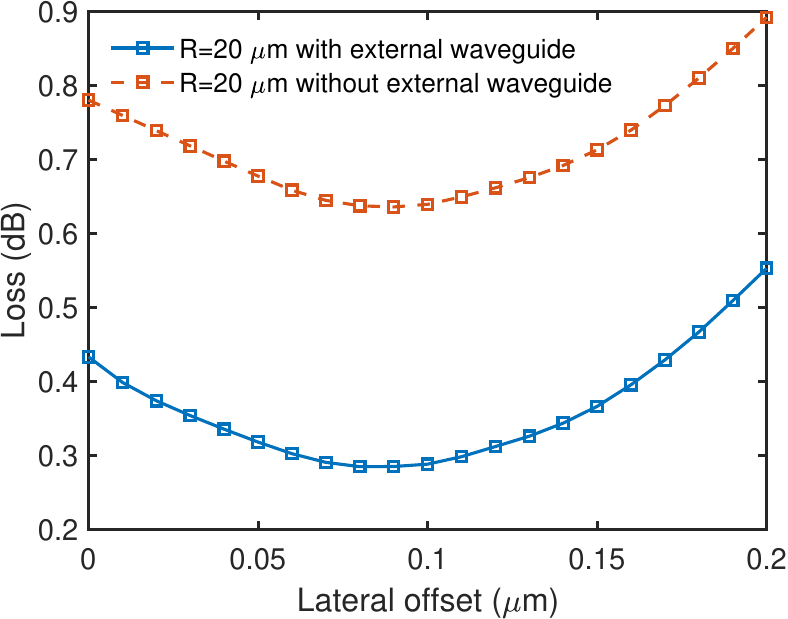}
  \end{tabular}
\caption{TE transmission loss as a function of the lateral offset with and without the asymmetric coupled external waveguide.  The results in the left plot correspond to $R=15$ $\upmu$m and those in the right plot to $R=20$ $\upmu$m.\label{fig::loffTE}}
\end{figure}

Figure \ref{fig::loffTE} displays the results obtained with the EME method for the conventional geometry of Figure \ref{fig::geometry} (left), where the impact of the straight--bent waveguide discontinuity was minimized using the lateral offset technique. Results for $R=15$ $\upmu$m and $R=20$ $\upmu$m are, respectively, displayed in the left and right plots.  Calculations were performed for a single curved waveguide and the case of the asymmetric coupled waveguides.

For a 180$^\circ$ bend, when the main waveguide without the external coupled section at $R=15$ $\upmu$m was considered, a minimal transmission loss of $1.75$ dB was obtained for a lateral offset of $0.12$ $\upmu$m.  This attenuation value includes both the contribution from the curvature discontinuities and that of the propagation along the bent section.  The design of the asymmetric coupled constant curvature arc was based on analysis of the modal properties of the TE quasi-mode.  The imaginary part of the effective index was found to have a minimum at $s=1.3$ $\upmu$m and $w_e=0.6$ $\upmu$m.  The addition of the external waveguide permitted reducing the total minimal radiation loss to  $0.82$ dB for an offset of $0.11$ $\upmu$m.  The fact that the optimal offset was shifted to a smaller value for the asymmetric coupled waveguide geometry evidences how the addition of a properly designed external waveguide contributed to improving the field confinement in the main waveguide, reducing the outwards shift of the optical radiation propagating in the bent section.  Therefore, the introduction of the asymmetric coupled waveguide provided a reduction factor of $0.83$  of the radiation loss of a 180$^\circ$ bend.  This figure amounts to half the round-trip loss if such a bend is part of a racetrack micro-resonator and results in a significant increase in the unloaded resonator Q factor \cite{Q,Qexp}. 

When the waveguide radius was increased to a value of $R=20$ $\upmu$m, a minimal loss of $0.64$ dB in the absence of the external waveguide was obtained for an offset of $0.09$ $\upmu$m.  The optimal configuration for the external waveguide section in the asymmetric coupled case is given by $w_e=1.7$ $\upmu$m and $s=0.6$ $\upmu$m.  In this case, a minimum loss of $0.29$ dB was obtained with a lateral offset of $0.08$ $\upmu$m.  We can observe the reduction in the optimal values of the lateral offsets for the larger value of $R$ associated with a less severe curvature-induced distortion of the modal field.  Again, a smaller optimal value of lateral offset when the external waveguide was present evidences its contribution to the confinement of the optical field in the main waveguide.  In this case, the radiation loss for a 180$^\circ$ bend was reduced by a factor of 0.92.

\begin{figure}[h]
	\begin{tabular}{cc}
		\includegraphics[width=6.5cm]{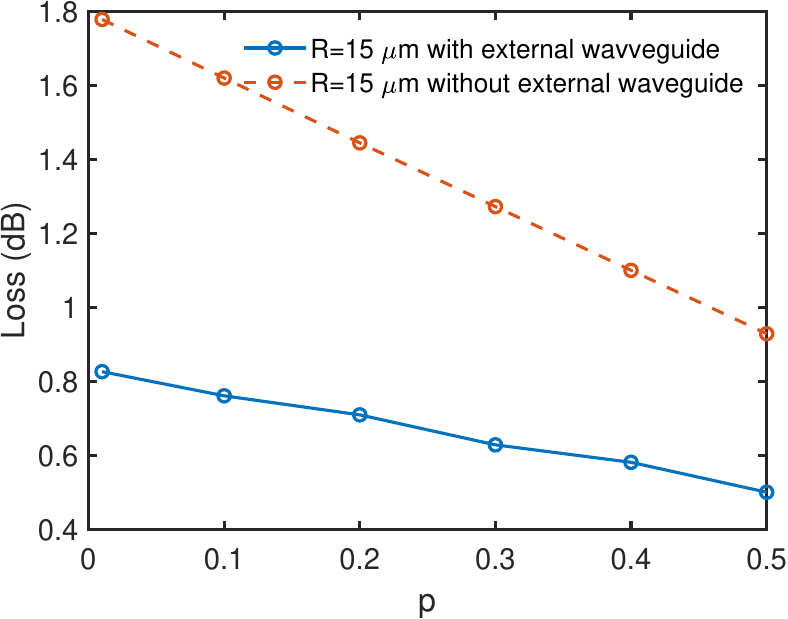}&
		\includegraphics[width=6.5 cm]{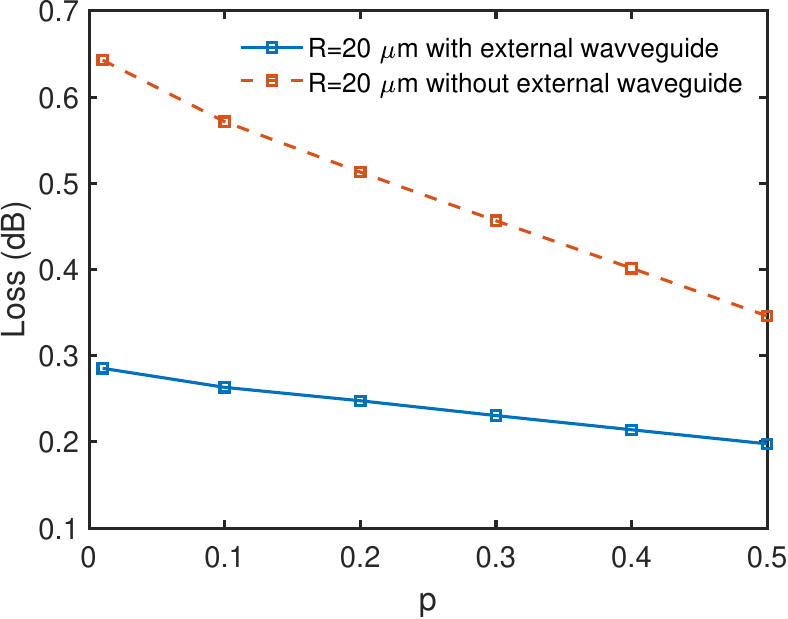}
	\end{tabular}
\caption{TE transmission loss with a partial Euler bend and the asymmetric coupled external waveguide in the constant curvature section.  The results for $R=15$ $\upmu$m and $R=20$ $\upmu$m are shown in the left and right plots, respectively. \label{fig::pTE}}
\end{figure}

Figure \ref{fig::pTE} (left) displays the total radiation loss for a 180$^\circ$ partial Euler bend as a function of the parameter $p$. Results for ${R=15\,\,\upmu\text{m}}$ and  ${R=20\,\,\upmu\text{m}}$ at the constant curvature section are shown in the left and right plots, respectively.   The results with and without the asymmetric coupled geometry in this region are both shown, demonstrating the significant loss reduction provided by an adequate external waveguide.   When this element was included, the total loss for ${R=15\,\,\upmu\text{m}}$ was $0.83$ dB  for very small lengths of the Euler section, $p=0.01$.  This figure is  very close to the value obtained with an abrupt lateral offset implementation.  A brusque lateral offset geometry cannot be accurately reproduced by conventional fabrication methods, which will create a smoothed version with a given degree of uncertainty.  One can effectively replace this design with a short (small $p$) partial Euler section, large enough so as to be reproduced with high fidelity in the fabricated device, and with very similar performance to that of the lateral offset alternative. The results for ${R=20\,\,\upmu\text{m}}$ at the constant curvature section are displayed in the right plot in Figure \ref{fig::pTE}. The smaller curvature resulted in a lower loss but at the cost of an increase in the surface used.  In this case, a total attenuation of $0.29$ dB was obtained with the inclusion of the external waveguide for $p=0.01$, which coincides with the theoretical value for a sharp geometry implementing an optimal lateral offset design. Both plots of Figure \ref{fig::pTE} show a similar gradual decrease in the total loss of the partial Euler bend as the value of $p$ grows.

The values of $p$ in the numerical survey do not cover the full range between $0$ and $1$, in spite of the clear improvement trend with increasing $p$ displayed  by the results in Figure \ref{fig::pTE}.  The reason for this is that the reduction in radiation loss as $p$ grew was obtained at the cost of a consequent enlargement of the device surface.  The limit value of $p=1$ corresponds to a full Euler bend without a coupled asymmetric waveguide section, which has a maximal surface for a fixed value of $R$.  The trade-off existing between radiation loss and device surface is analyzed in the next section.  We find that, for a given loss target, the introduction of the proposed radiation-reduction geometry permits obtaining the same performance level with a smaller footprint by simultaneously reducing the value of $p$ and making $R$ larger.  Therefore, this study was limited to values of $p$ up to $p=0.5$, effectively setting a lower bound on the contribution from the asymmetric coupled waveguide sector.

\subsection{Device Footprint}\label{sec::footprint}

One of the main advantages of asymmetric coupled waveguide geometries is that they permit reducing the device surface.  Therefore, when evaluating the performance of the various bend geometries, the total area spanned by the device has to be carefully taken into account.

\begin{figure}[h]
  \includegraphics[width=6.5 cm]{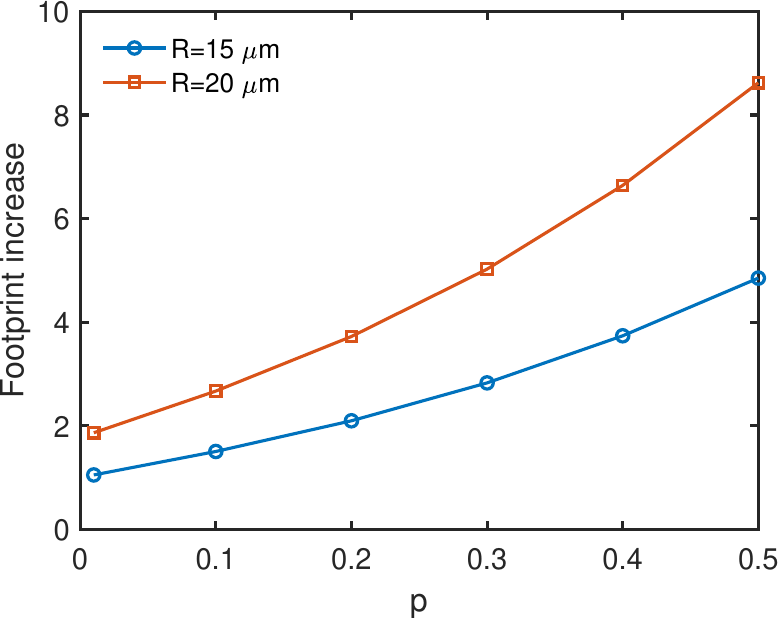}
  
\caption{Surface of the $R=15$ $\upmu$m and $R=20$ $\upmu$m polarizers with partial Euler bends relative to the footprint of a lateral offset $R=15$ $\upmu$m  device as a function of $p$.\label{fig::footprint}}
\end{figure}

The results in Figure \ref{fig::pTE} show, for each value of $R$ and constant bent section design, a steady decrease in the total attenuation as the parameter $p$ increases.  Figure \ref{fig::footprint} displays the corresponding factors for the increment in the bend surface, relative to the smallest value obtained for the lateral offset implementation of a ${R=15\,\,\upmu\text{m}}$ 180$^\circ$ bend, as a function of $p$ for the two values of $R$ addressed in this work.  At $p=0.2$, the device surface for ${R=15\,\,\upmu\text{m}}$ became larger than that of the smallest $p=0.01$ ${R=20\,\,\upmu\text{m}}$ design, while the loss level was still larger in the ${R=15\,\,\upmu\text{m}}$, $p=0.2$ case.  Therefore, the compromise existing between device footprint requirements and tolerable transmission loss must be carefully evaluated in the design of partial Euler bend geometries incorporating asymmetric coupled waveguides.  For instance, once the value of $R$ has been set, $p$ can be used to trim the total loss within certain device size restrictions.  If the size limit is exceeded, a solution with a larger value of $R$ and smaller $p$, having a stronger contribution from the asymmetric coupled waveguide section, should be sought.

\subsection{Polarizers}

As in the case of the reduced curvature loss discussed above, when the geometry in Figure~\ref{fig::geometry} is used for the selective attenuation of a given polarization, the device design starts with the analysis of the modal properties of the constant curvature section, i.e., region III.  The polarizer performance is defined by two parameters: the polarization extinction ratio (PER), which is the ratio between the attenuation of the undesired TM polarization and that of the TE polarization component, and the insertion loss (IL) characterizing the attenuation of the desired TE component.  We seek operation points in the design space defined by parameters $w_e$ and $s$ that simultaneously provide a large value of PER and small IL.  A good compromise is obtained by the optimization of the quotient $n_{i,TM}/n_{i,TE}$, where $n_i$ is the imaginary part of the complex refractive index of the quasi-modes in the bent sections.  This figure of merit permits simultaneously balancing a large value of PER at small IL \cite{POLexp}.   

\begin{figure}[h]
  \centering
  \begin{tabular}{cc}
  \includegraphics[width=6.5 cm]{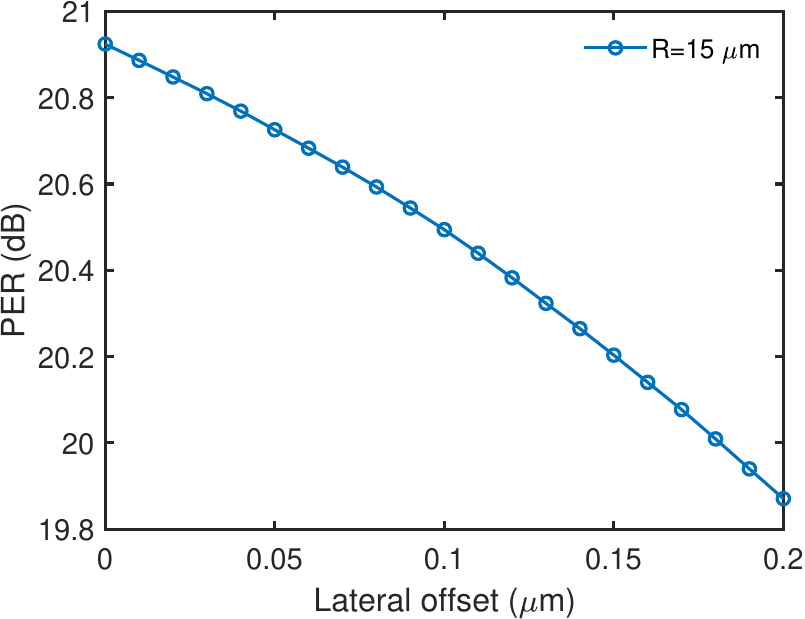}&
  \includegraphics[width=6.5 cm]{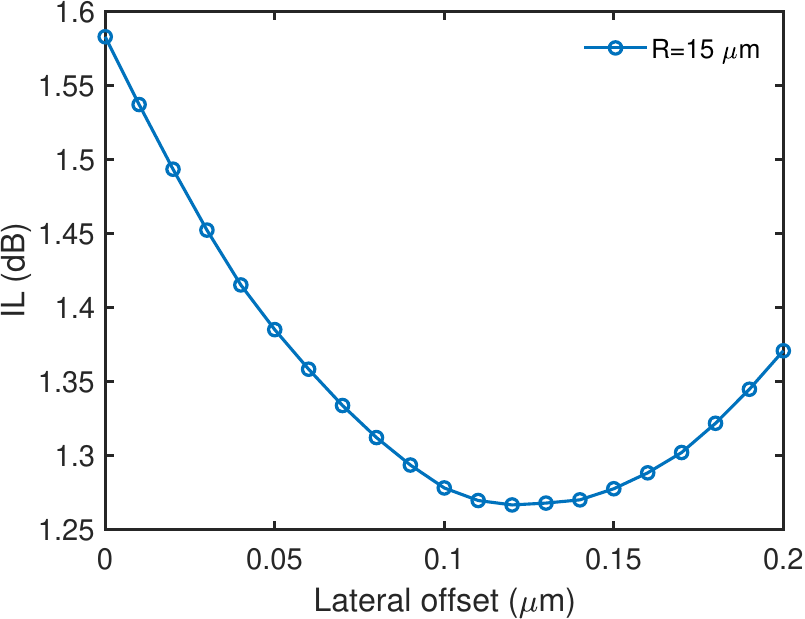}\\
   \includegraphics[width=6.5 cm]{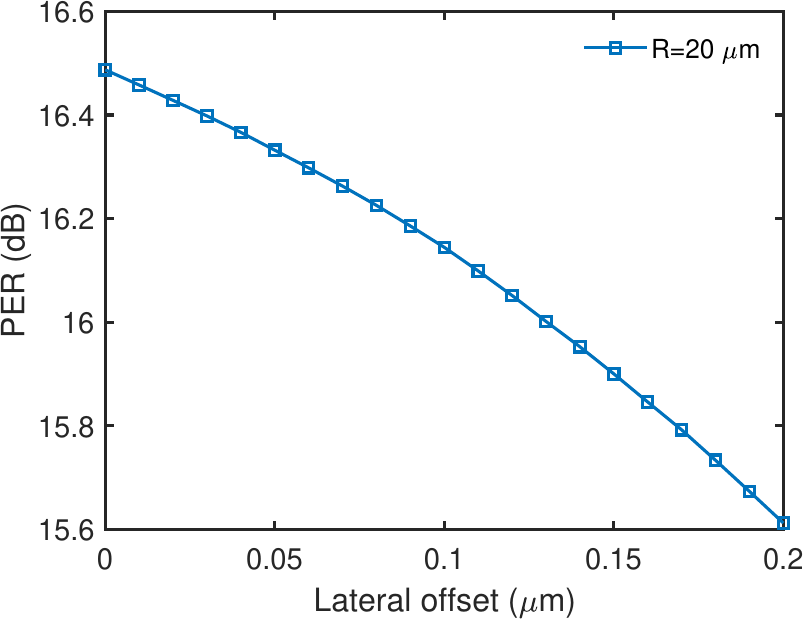}&
  \includegraphics[width=6.5 cm]{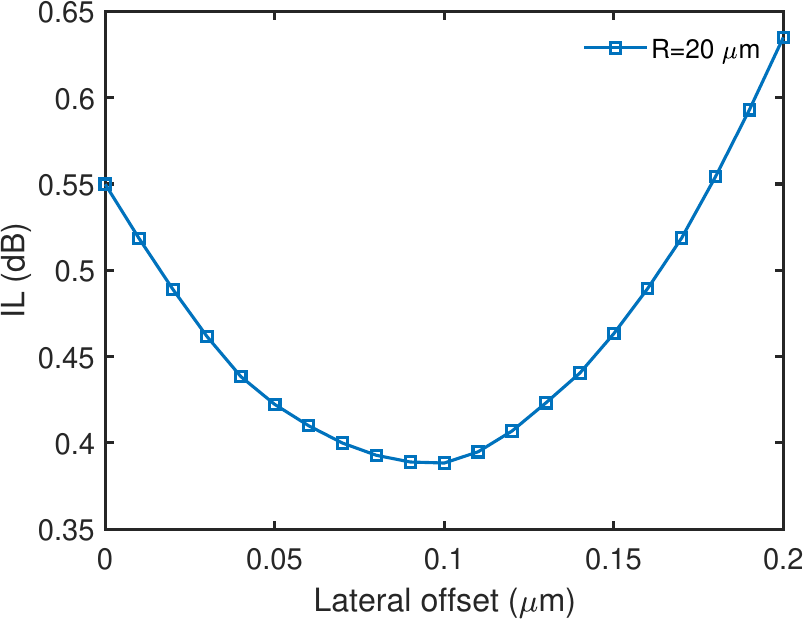}
  \end{tabular}
\caption{Calculated values of PER (\textbf{left}) and IL (\textbf{right}) as a function of the lateral offset for the device geometry depicted in Figure \ref{fig::geometry}.  Plots in the top and bottom rows correspond, respectively, to \mbox{$R=15$ $\upmu$m} and $R=20$ $\upmu$m.\label{fig::loff}}
\end{figure}

Two values of $R$, $15$ and $20$ $\upmu$m, were analyzed for the constant curvature sections.  The optimization was performed using the FIMMWAVE mode solver, yielding parameters $w_e=0.5$ $\upmu$m and $s=1.8$ $\upmu$m for $R=15$ $\upmu$m and $w_e=0.5$ $\upmu$m and $s=2.1$ $\upmu$m for $R=20$ $\upmu$m. Figure \ref{fig::loff} displays the values of PER (left) and IL (right) for the geometry of Figure \ref{fig::geometry} as a function of the lateral offset calculated using FIMMPROP.  The results for $R=15$ $\upmu$m and $R=20$ $\upmu$m are shown in the top and bottom rows, respectively.  The values obtained for the lateral offset that minimize the loss for the desired TE polarization were in good agreement with those previously obtained maximizing the overlap of the field distributions in the straight and curved waveguide \cite{POL}.  It can be observed how the lateral offset also had a certain polarization-dependent behavior and contributed in a milder way to the total value of the PER.  The main drawback of the design approach based on the lateral offset is that the actual geometry will differ from the original design, due to fabrication limitations.  The resolution restraints in lithographic techniques will smooth the sharp features of the ideal lateral offset.  Even though the responses measured on fabricated devices were found close to those of the ideal designs \cite{POLexp,Qexp}, such uncertainty is undesirable in high-accuracy PIC design.

\begin{figure}[h]
  \centering
  \begin{tabular}{cc}
  \includegraphics[width=6.5 cm]{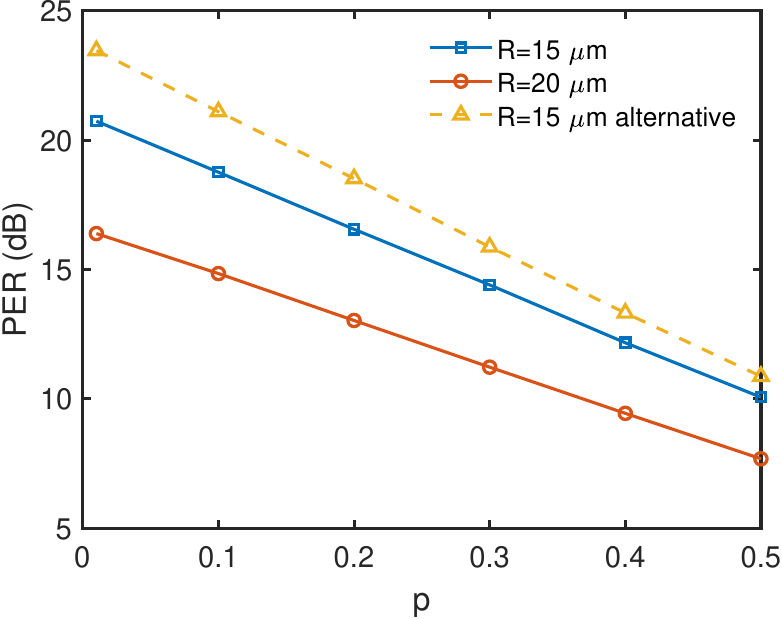}&
  \includegraphics[width=6.5 cm]{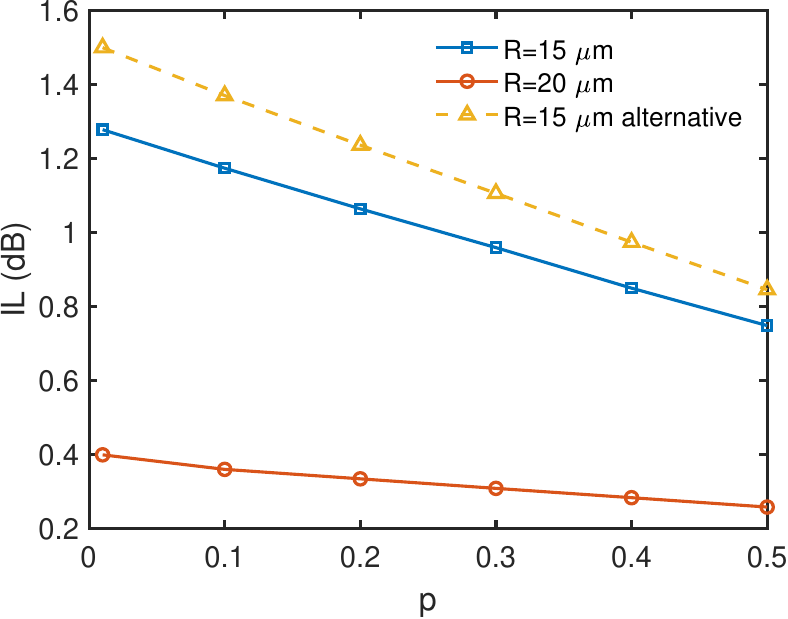}
  \end{tabular}
  \caption{Calculated extinction ratio {(\textbf{left})} 
 and insertion loss {(\textbf{right})} as a function of the bend parameter $p$ for $R=15\,\upmu$m and $R=20\, \upmu$m at $\lambda_0=1550\,$nm.  Dashed lines correspond to an alternative $R=15\,\upmu$m design not based on the optimization of $n_{i,TM}/n_{i,TE}$.\label{fig::pol}}
\end{figure}

Figure \ref{fig::pol} displays the values of PER and IL calculated for structures incorporating partial Euler bends, as described in Figure \ref{fig::geometry},  for values of $p$ varying between $0.01$ and $0.5$.  The results show an approximately linear dependence of both PER and IL with $p$.  For very small $p=0.01$, the results of PER and IL were very close to the abrupt lateral offset results shown in Figure \ref{fig::loff} at the lateral offset values corresponding to the IL minima.  Therefore, the use of partial Euler bends permits expanding, in an optimal way, the mode matching sections to a smooth geometry that can be reproduced with high fidelity in the fabrication process.

The fact that PER and IL can be simultaneously trimmed by introducing partial Euler bends also introduces more flexibility into the design process, where several sections like the one depicted in Figure \ref{fig::geometry} can be combined to achieve a given device performance.  The existence of a new design parameter $p$ can be exploited to explore designs alternative to the aforementioned condition given by the maximization of the ratio $n_{i,TM}/n_{i,TE}$.  This is illustrated in the results plotted with a dashed trace in Figure \ref{fig::pol}, providing larger values of both IL and PER.

When exploring polarizer designs based on partial Euler bend geometries, it is  also important to bear in mind the impact of the value of $p$ on the device footprint.  This effect is displayed in Figure \ref{fig::footprint}, where the surfaces of $R=15$ $\upmu$m and $R=20$ $\upmu$m polarizers with different values of $p$ are compared with the reference value of a lateral offset \mbox{$R=15$ $\upmu$m} polarizer. The values of the design parameters, $R$ and $p$, must be carefully chosen in accordance with the requirements of PER and IL, as well as the device footprint restrictions. 

\vspace{-6pt}

\section{Discussion}

The incorporation of smooth transitions shaped as partial Euler bends in asymmetric coupled waveguides was analyzed.  Two different scenarios were addressed: designs for the reduction of total radiation loss of the desired TE polarization in curved geometries, and integrated polarizers designed to reject the undesired TM polarization, while minimizing the loss of the orthogonal TE field component.

In both scenarios, the use of short Euler sections permitted approaching the performance levels of the previously analyzed and experimentally demonstrated geometries based on abrupt transitions at the straight--bent sections incorporating lateral offsets.  Such sharp features are beyond the resolution of the fabrication process and become smoothed out in an uncontrollable manner in the produced PICs.  Therefore, short clothoid transitions can be effectively incorporated as a replacement for lateral offsets, to produce designs that can be reproduced in the PICs with high accuracy at the same theoretical performance level.     

The analysis of radiation loss reduction geometries showed that a significant improvement in transmission loss can be obtained in partial Euler bends by incorporating asymmetric coupled waveguides into the constant curvature sections, thus permitting a reduction in the radii of curvature for a given level of loss and a consequent shrinkage of the occupied surface. 

The introduction of a new parameter defining the angular ratio corresponding to the adiabatic section of linearly varying curvature provides a new degree of freedom in device design, but a careful evaluation of the trade-off existing between the device performance and footprint must be taken into account. 
\section*{Acknowledgement}

This work was funded by the Spanish Ministerio de Ciencia e Innovación (MCIN), project PID2020-119418GB-I00, and the European Union NextGenerationEU (PRTRC17.I1) and the Consejería de Educación, Junta de Castilla y Le\'on, through QCAYLE project.


\begin{thebibliography}{999}

  %
  %
\bibitem{zhang}
  Zhang, C.; Tran, M.A.; Dorche, A.E.; Shen, Y.; Shen, B.; Asawa, K.; Kim, G.; Kim, N.; Levison, F.; Bowers, J.E.; Komljenovic, T. Integrated photonics beyond communications. {\em Appl. Phys. Lett.} {\bf 2023}, {\em 123}, 230501.


\bibitem{wuJ}
  Wu, J.; Yue, G.; Chen W.; Xing, Z.; Wang, J.; Wong, W.R.; Cheng, Z.; Set, S.Y.; Murugan, G.S.; Wang, X.; Liu, T.  On-Chip Optical Gas Sensors Based on Group-IV Materials. {\em ACS Photonics} {\bf 2020}, {\em 7}, 2923--2940.

\bibitem{wang}
  Wang, J.; Dong, J. Optical Waveguides and Integrated Optical Devices for Medical Diagnosis, Health Monitoring and Light Therapies. {\em Sensors} {\bf 2020}, {\em 20}, 3981.
  
\bibitem{chang}
  Chang, L.; Liu, S.; Bowers, J.E. Integrated optical frequency comb technologies. {\em Nat. Photonics} {\bf 2022}, {\em 16}, 95--108.

\bibitem{luo}
 Luo, W.;  Cao, L.; Shi, Y.; Wan, L.; Zhang, H.; Li, S.; Chen, G.;  Li, Y.;  Li, S.; Wang, Y.; Sun, S.;  Karim, M.F.; Cai, H.;  Kwek, C.L.; Liu, A.Q.  Recent progress in quantum photonic chips for quantum communication and internet. {\em Light Sci. Appl.} {\bf 2023}, {\em 12}, 175.
  
  
\bibitem{siew}
  Siew, S.Y.; Li, B.; Gao, F.; Zheng, H.Y.; Zhang, W.; Guo, P.; Xie, S.W.; Dong, L.; Luo, L.W.; Li, C.; Luo, X.; Lo, G.-Q. Review of Silicon Photonics Technology and Platform Development.  {\em J. Light. Technol.} {\bf 2021}, {\em 39}, 4374--4389.

\bibitem{sharma}
  Sharma, T.; Wang, J.; Kaushik, B.K.; Cheng, Z.; Kumar, R.; Wei, Z.; Li, X. Review of Recent Progress on Silicon Nitride-Based Photonic Integrated Circuits. {\em IEEE Access} {\bf 2020}, {\em 8}, 195436--195446.

  %
  %

  \bibitem{radiation}
  Chamorro-Posada, P. Radiation in bent asymmetric coupled waveguides. {\em Appl. Opt.} {\bf 2019}, {\em 58}, 4450--4457.

  %
  %

\bibitem{wu}
Wu, H.; Li, C.; Song, L.; Tsang, H-K.; Bowers, J.E.; Dai, D. Ultra-Sharp Multimode Waveguide Bends with Subwavelength Gratings. {\em Laser Photonics Rev.} {\bf 2019}, {\em 13}, 1800119.

\bibitem{harjanne}
  Harjanne, M.;  Aalto, T. Design of tight bends in silicon-on-insulator ridge waveguides. {\em Phys. Scr.} {\bf 2004}, {\em T114}, 209--212.

\bibitem{sole}
Solehmainen, K.; Aalto, T.; Dekker, J.; Kapulainen, M.; Harjanne, M.; Heimala, P. Development of multi-step processing in silicon-on-insulator for optical waveguide applications.  {\em J. Opt. A Pure Appl. Opt.} {\bf 2006}, {\em 8}, S455--S460.
 


  %
  %
  \bibitem{melloni2003} Melloni, A.; Monguzzi, P.; Costa, R.; Martinelli, M. Design of curved waveguides: The matched bend. {\em J. Opt. Soc. Am. A} {\bf 2003}, {\em 20}, 130--137.

  %
  %

\bibitem{neumann} Nuemann, E.-G. Curved dielectric optical waveguides with reduced transition losses. {\em IEE Proc. } {\bf 1982}, {\em 129 Pt. H}, 278--280.



\bibitem{kitoh} Kitoh, T.; Takato, N.; Yasu, M.; Kawachi, M. Bending loss reduction in silica-based waveguides by using lateral offsets. {\em J. Light. Technol.} {\bf 1995}, {\em 13}, 555--562.

  %
  %
\bibitem{ladouceur} Ladouceur, F.; Labeye, P. A New General Approach to Optical Waveguide Path Design. {\em J. Light. Technol.} {\bf 1995}, {\em 13}, 481--492.
  \bibitem{yi} Yi, D.; Zhang, Y.; Tsang, H.K. Optimal Bezier curve transition for low-loss ultra-compact S-bends. {\em Opt. Lett.} {\bf 2021}, {\em 46}, 876--879.

  \bibitem{song} Song, J.H.; Kongnyuy, T.D.; Stassen, A.; Mukund, V.; Rottenberg X.  Adiabatically Bent Waveguides on Silicon Nitride Photonics for Compact and Dense Footprints. {\em IEEE Photonics Technol. Lett.} {\bf 2016}, {\em 28}, 2164--2167.
\bibitem{song2020} Song, J.H.; Kongnyuy, T.D.; De Heyn, P.; Lardenois, S.; Roelof, J.; Rottenberg, X. Low-Loss Waveguide Bends by Advanced Shape for Photonic Integrated Circuits.  {\em J. Light. Technol.} {\bf 2020}, {\em 38}, 3273--3279.
    
  %
  %
\bibitem{liu} Liu, P.-L.; Li, B.-J., Cressman, P.J.; Debesis, J.R.; Stoller, S. Comparison of Measurede Losses of $\text{Ti:LiNbO}_3$ Channel Waveguide Bends. {\em IEEE Photonics Technol. Lett.} {\bf 1991}, {\em 3}, 755--756.
  %
  %
\bibitem{mustieles} Mustieles, F.J.; Ballesteros, E.; Baquero, P. Theoretical S-Bend Profile for Optimization of Optical Waveguide Radiation Losses. {\em IEEE Photonics Technol. Lett.} {\bf 1993}, {\em 5}, 551--553.
  \bibitem{bogaerts} Bogaerts, W.; Selvaraja, S.K. Compact Single-Mode Silicon Hybrid Rib/Strip Waveguide With Adiabatic Bends.  {\em IEEE Photonics J.} {\bf 2011}, {\em 3}, 422--432.
  %
  %
  \bibitem{cherchi} Cherchi, M.; Ylinen, S.; Harjanne, M.; Kapulainen, M.; Aalto, T. Dramatic size reduction of waveguide bends on a micron-scale silicon photonic platform.  {\em Opt. Express} {\bf 2013}, {\em 21}, 17814--17823.
  \bibitem{jiang} Jiang, X.; Wu, H.; Dai, D.  Low-loss and low-crosstalk multimode waveguide bend on silicon.  {\em Opt. Express} {\bf 2018}, {\em 13}, 17680--17689.
\bibitem{fujisawa} Fujisawa, T.; Makino, S.; Sato, T.; Saitoh, K. Low-loss, compact, and fabrication-tolerant Si-wire $90\deg$ waveguide bend using clothoid and normal curves for large scale photonic integrated circuts. {\em Opt. Express} {\bf 2017}, {\em 25}, 9150--9159. 
\bibitem{vogelbacher} Vogelbacher, F.; Nevlcasil, S.; Sagmeister, M.; Kraft, J.; Unterrainer, K.; Hainberger, R. Analysis of silicon nitride partial Euler waveguide bends. {\em Opt. Express} {\bf 2019}, {\em 27}, 31394--31406.

  %
  %
\bibitem{sun} Sun, T.; Xia, M. Low loss modified Bezier bend waveguide.  {\em Opt. Express} {\bf 2022}, {\em 30}, 452580.
  %
  %
\bibitem{zhang2023} Zhang, L.; Chen, J.; Ma, W.; Chen, G.; Li, R.; Li, W.; An, J.; Zhang, J.; Wang, Y.; Gou, G.; Liu, C.; Qi, Z.; Xue, N. Low-loss, ultracompact $n$-adjustable waveguide bends for photonic integrated circuits. {\em Opt. Express} {\bf 2023}, {\em 31}, 2792--2806.
  %
  %
\bibitem{chen2012} Chen, T.; Lee, H.; Li, J.; Vahala, K.J. A general design algorithm for low optical loss adiabatic connections in waveguides. {\em Opt. Express} {\bf 2012}, {\em 20}, 22819–22829.

\bibitem{gabrielli} Gabrielli, L.H.;  Liu, D.; Johnson, S.G.; Lipson, M. On-chip transformation optics for multimode waveguide bends. {\em Nat. Commun.} {\bf 2012} {\em 3}, 1217.
  

\bibitem{yu2018} Yu, Z.; Ma, Y.; Sun, X. Photonic welding points for arbitrary on-chip optical interconnects. {\em Nanophotonics} {\bf 2018}, {\em 7}, 1679–1686.

   

\bibitem{liu2018} Liu, Y.; Sun, W.; Xie, H.; Zhang, N.; Xu, K.; Yao, Y.; Xiao, S.; Song, Q. Very sharp adiabatic bends based on an inverse design. {\em Opt. Lett.} {\bf 2018}, {\em 43}, 2482–2485.

  
\bibitem{bahadori} Bahadori, M.N.M.; Cheng, Q.X.; Bergman, K. Universal design of waveguide bends in silicon-on-insulator photonics platform.  {\em J. Light. Technol.} {\bf 2019}, {\em 37},  3044–3054.
 
\bibitem{li2020} Li, Z.; Li, G.; Huang, J.; Zhang,, Z.; Yang, J.; Yang, C.; Qian, Y.; Xu, W.; Huang, H. Ultra-compact high efficiency and low crosstalk optical interconnection structures based on inverse designed nanophotonic elements. {\em Sci. Rep.} {\bf 2020}, {\em 10}, 11993.

\bibitem{yu2020} Yu, Z.; Sun, X.. Inverse-designed photonic jumpers with ultracompact size and ultralow loss. {\em  J. Light. Technol.} {\bf 2020}, {\em 38}, 6623–6628.


  %
  %


  %
  \bibitem{Q}
  Chamorro-Posada, P. Q-Enhanced racetrack microresonators. {\em Optics Commun.} {\bf 2017}, {\em 387}, 70--78.
 %
\bibitem{Qexp}
  Chamorro-Posada, P.; Baños, R. Design and Characterization of Q-Enhanced Silicon Nitride Racetrack Microresonators. {\em J. Light. Technol.} {\bf 2021}, 2917--2923.
  
  \bibitem{biosensors} Chamorro-Posada, P. Asymmetric Concentric Microring Resonator Label-Free Biosensors. {\em Photonics} {\bf 2022}, {\em 9}, 9010027
  %
  \bibitem{POL}
  Chamorro-Posada, P. Ultracompact integrated polarizers using bent asymmetric coupled waveguides. {\em Opt. Lett.} {\bf 2019}, {\em 44}, 2040--2043.
%
\bibitem{POLexp}
  Chamorro-Posada, P. Design and characterization of silicon nitride ultracompact integrated polarizers using bent asymmetric coupled waveguides. {\em Opt. Lett.} {\bf 2021}, {\em 46}, 609--612.
  %


\bibitem{abramowitz}
  Abramowitz, M. \textit{Handbook of Mathematical Functions, with Formulas, Graphs, and Mathematical Tables}; Dover Publications: Mineola, NY, 
 USA, 1974; p. 301.



\bibitem{krause}
  Krause, M. Finite-difference mode solver for curved waveguides with angled and curved dielectric interfaces. {\em J. Light. Technol.} {\bf 2009}, {\em 29}, 691--699.

\bibitem{wgms3d}
wgms3d-Full-Vectorial Waveguide Mode Solver. Available online: \url{http://www.soundtracker.org/raw/wgms3d/}  {(accessed on 27 February 2024).} 


  \bibitem{hiremath}
Hiremath, K.R.; Hammer, M.; Stoffer, R.; Prkna, L.; Ctyroky, J., Analytic approach to dielectric optical bent slab waveguides. {\em Opt. Quantum Electron.} {\bf 2005}, {\em 37}, 37--61.
  %
  %

\bibitem{fimmprop}
  Gallagher, D.F.G.; Felici, T.P. Eigenmode expansion methods for simulation of optical propagation in photonics: Pros and cons. {\em Proc. SPIE} {\bf 2003}, {\em 4897},  69--89.
  






\bibitem{huang}
 Huang, W.P.; Xu, C.-L. Simulation of Three-Dimensional Optical Waveguides by a Full-Vector Beam Propagation Method. {\em IEEE J. Quantum Electron.} {\bf 1993}, {\em 29}, 2639--2649.

 
\bibitem{lui}
Lui, W.; Xu, C.-L.; Hirono, T.; Yokoyama, K.; Huang, W.P. Full-vectorial wave propagation in semiconductor optical bending waveguides and equivalent straight waveguide approximations.   {\em J. Light. Technol.} {\bf 1998}, {\em 16}, 910--914.

  
\bibitem{fdtd}
  Yee, K, Numerical solution of initial boundary value problems involving Maxwell's equations in isotropic media. {\em IEEE Trans. Antennas Propag.} {\bf 1966}, {\em 14}, 302--307.



\end{thebibliography}
\end{document}